\documentclass[10pt,aps,prx,twocolumn,notitlepage,showpacs,superscriptaddress, 
]{revtex4-1}
\usepackage{graphicx}
\usepackage[english]{babel}
\usepackage{amsthm}
\usepackage{amssymb,amsmath,bm}
\usepackage{footnote}
\usepackage[colorlinks, linkcolor=blue,anchorcolor=blue,citecolor=blue,urlcolor=blue]{hyperref}
\newcommand{\be}{\begin{equation}}
\newcommand{\ee}{\end{equation}}

\renewcommand{\b}[1]{{\boldsymbol{#1}}}

\newcommand{\sgn}{\mathop{\mathrm{sgn}}}

\begin{document}
\title{Introduction to dimensional reduction of fermions}
\author{Joel Hutchinson}
\affiliation{Department of Physics, University of Basel, Klingelbergstrasse 82, CH-4056 Basel, Switzerland} 

\author{Dmitry Miserev}
\affiliation{Department of Physics, University of Basel, Klingelbergstrasse 82, CH-4056 Basel, Switzerland} 

\author{Daniel Loss}
\affiliation{Department of Physics, University of Basel, Klingelbergstrasse 82, CH-4056 Basel, Switzerland} 

\author{Jelena Klinovaja}
\affiliation{Department of Physics, University of Basel, Klingelbergstrasse 82, CH-4056 Basel, Switzerland} 
\date{\today}
\begin{abstract}
We present a comprehensive pedagogical introduction to the dimensional reduction protocol (DRP), a versatile framework for analyzing instabilities and critical points in interacting fermionic systems. The DRP simplifies the study of many-body problems by systematically reducing their effective spatial dimension while retaining essential physics. This method works for electron gases in a diverse array of settings: in any number of spatial dimensions, in the presence of Zeeman fields, with spin-orbit coupling, including repulsive or attractive interactions. 
Focusing on two-point correlation functions, the DRP identifies a minimal subspace relevant for capturing analytic properties, facilitating efficient computation of critical phenomena in electronic systems. This work outlines the assumptions, proof, and applications of the DRP, emphasizing its simplicity and broad applicability for future studies in correlated electron physics.
\end{abstract}
\maketitle

\section{Introduction}

The purpose of this paper is to provide a pedagogical outline of a method that has been used in a series of recent works to determine properties of instabilities and critical points in a variety of interacting electronic systems~\cite{miserev2021a, miserev2022, miserev2023, hutchinson2024, miserev2024, miserev2025}. The method doe not discriminate with regards to the dimensionality of the electron gas. It accommodates Zeeman fields and spin-orbit coupling, single and multilayer heterostructures. Nor does it discriminate with respect to repulsive or attractive interactions, allowing one to study charge/spin instabilities or superconducting instabilities with equal ease. These are only the applications considered thus far and many more possibilities lay open for future work. It rests on only a few well-defined assumptions, and its simplicity makes it an essential tool for anyone studying correlation effects in fermionic systems. We will refer to this method as the dimensional reduction protocol (DRP).

\section{Heuristic motivation}
When employing the Feynman diagrammatic expansion of any quantity in an interacting many-body problem, one is faced with the difficulty that the complexity of a given diagram increases dramatically with the spatial dimension of the problem. In essence, the goal of the DRP is to systematically reduce the effective dimension of the system. To do so requires identifying the subspace of a $D$-dimensional problem that contains the physics essential to determine the analytic properties of a given function of interest.

\begin{figure}
  \includegraphics[width=0.99\columnwidth]{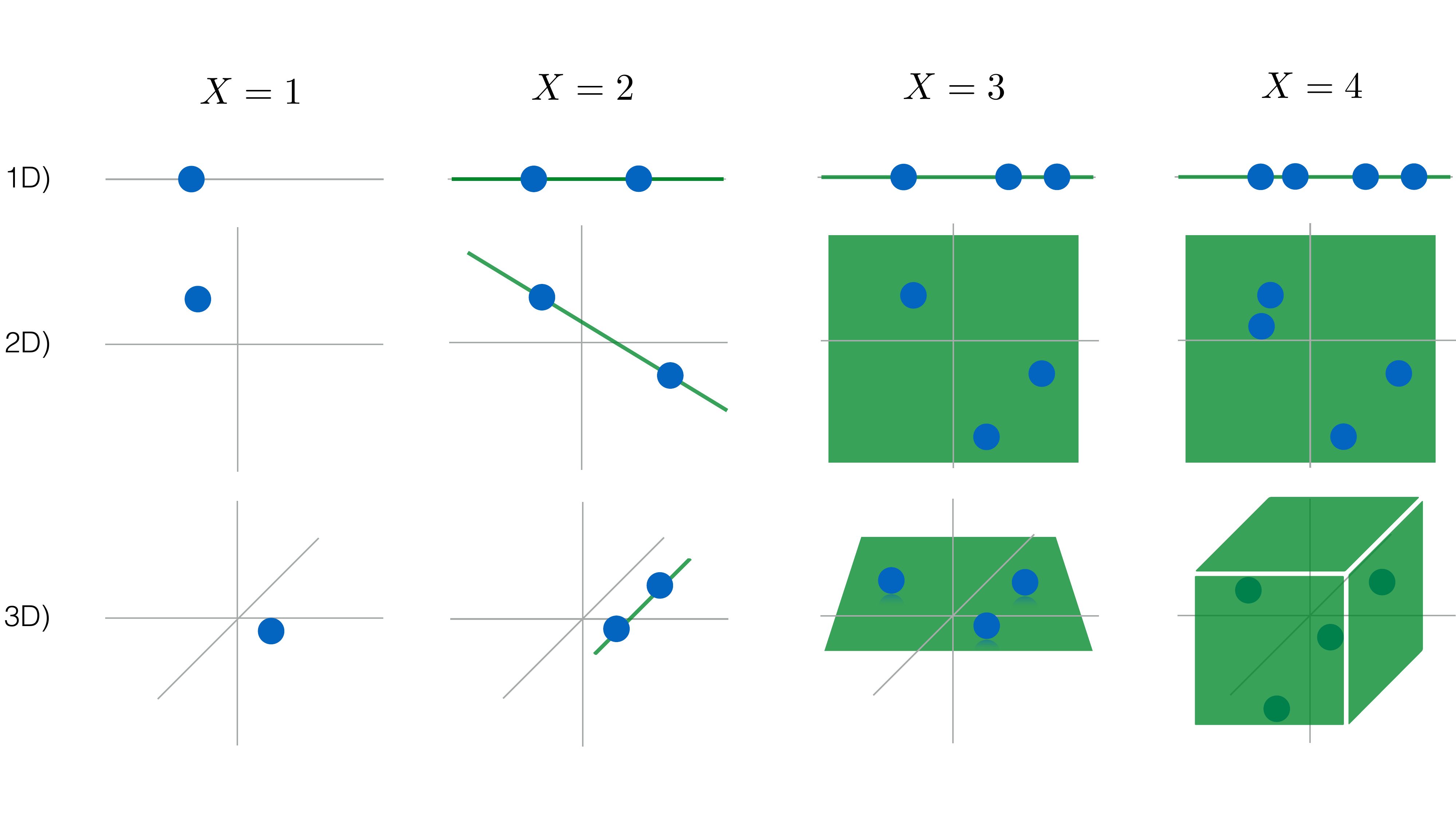}
  \caption{Minimal subspace (green) connecting $X$ external position vectors (blue) in different dimensions.}
  \label{fig:dim}
\end{figure}

One candidate subspace presents itself immediately upon considering Feynman diagrams in real space. Each external vertex comes with a vector that lives in $D$-dimensional space. For a diagram with $X$ external vertices, the minimal subspace connecting these vectors has dimension ${\rm min}(X-1,D)$ as shown in Fig.~\ref{fig:dim}. This subspace can be useful for reducing the computational difficulty of a the diagram provided $X\leq D$. In fact, we will be a bit more restrictive and focus on the case of correlation functions, for which $X=2$, and therefore define a natural one-dimensional subspace for all $D\geq1$.

Fortunately, two-point correlation functions 
are incredibly useful. They include many quantities connected to physical observables such as charge, spin and pair susceptibilities, and the self-energy. For systems with elastic scattering, these are functions of a single frequency and momentum $f(\b{q},\omega)$. To study criticality and phase transitions, we only care about the analytic behaviour of these functions within some small window of special $\omega$ and $\b{q}$ values (typically corresponding to scattering across a Fermi surface). In terms of the Fourier representation, $f(\omega,\b{q})=\int dte^{i\omega \tau}\int d^Dre^{-i\b{q}\cdot\b{r}}f(\b{r},\tau)$, this means we don't care about the low harmonics that lead to broad changes and constant shifts in reciprocal space. All of which is to say that we need an approximation framework that accurately captures the \emph{asymptotic} correlation function. Specifically, something valid for $r\gg\lambda_F$, $\tau\gg1/E_F$, where $E_F$ is the Fermi energy and $\lambda_F$ is the (largest) Fermi wavelength in the system. The situation is a bit reminiscent of the standard approach to potential-scattering problems in quantum mechanics where we throw away the near-field information of the scattering region in favour of a description in terms of free propagating waves with induced phase shifts from the scatterers. 

\section{Assumptions}\label{sec:assump}
The above considerations lead us to the semi-classical approximation:
\begin{itemize}
\item All bare Green's functions are replaced by their asymptotic values, 
\be
g_{\sigma\nu}(\tau, \boldsymbol{r})\approx\frac{1}{2 \pi\left(\lambda_F^\sigma r\right)^{\frac{D-1}{2}}} \sum_{s=\pm}\frac{e^{is\left(k_F^\sigma r-\vartheta\right)}}{is r-v_F^{\sigma\nu} \tau}.\label{eq:Greens}
\ee
\end{itemize}
Here $k_F^\sigma=2\pi/\lambda_F^\sigma$ is the Fermi wavenumber for a band indexed by $\sigma$, $v_F^{\sigma\nu}=\nu k_F^\sigma/m_*$ is the corresponding Fermi velocity, $\vartheta\equiv\pi(D-1)/4$ and $s=\pm1$ is referred to as the ``chirality index" for reasons that will become clear later. Note from the expression for $v_F$ that we have employed a second assumption,
\begin{itemize}
\item Fermi surfaces are isotropic but can be electron or hole-like as indicated by the sign $\nu=\pm1$ in $v_F$.
\end{itemize}
The coexistence of electron and hole Fermi surfaces can lead to pair-density waves~\cite{miserev2024}, and has a profound effect on the scattering and transport physics of the system~\cite{hutchinson2016, hutchinson2017, hutchinson2018}.

\begin{figure}
  \includegraphics[width=\columnwidth]{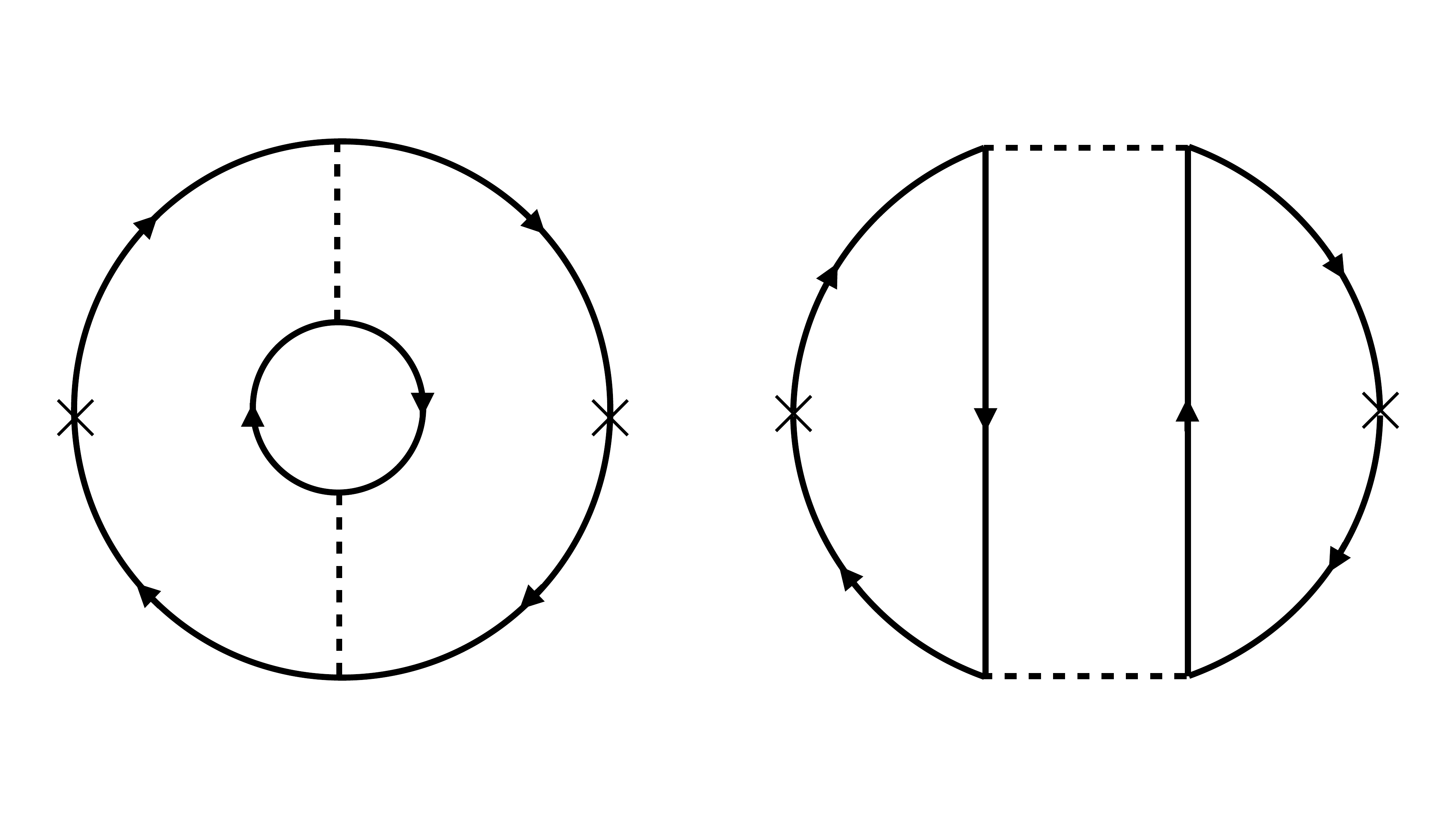}
  \caption{Examples of $N$-particle reducible diagrams with $N=2$ and two external vertices marked by crosses. The solid lines denote propagators and dashed lines interactions.}
  \label{fig:NPRed}
\end{figure}

To successfully use the DRP we require a third assumption.
\begin{itemize}
\item All diagrams under consideration are $N$-particle irreducible, where $N$ is the number of interaction lines. 
\end{itemize}
This means we will not consider diagrams that can be separated by cutting every interaction line. This would be the case for the dynamical screening and Aslamazov-Larkin diagrams as shown in Fig.~\ref{fig:NPRed}.
It should be noted that $N$-particle reducible diagrams are also amenable to the DRP, but each additional fermion loop adds one dimension to the minimal subspace~\cite{miserev2023}. Thus for two-dimensional systems, such diagrams do not benefit from the DRP~\footnote{We will next make the assumption that the interaction is of forward scattering type, thus additional fermion loops contain only small-momentum transfers and their contribution is always irrelevant. These reducible diagrams only become important if there are at least two vertices with large momentum transfer.}. 

In addition, we require that

\begin{itemize}
\item The interaction is forward-scattering type.
\end{itemize}
In other words, the interaction transfers momentum that is small compared to the Fermi momentum ($V(q=0)\gg V(q=2k_F)$). In Sec.~\ref{sec:back}, we will discuss this issue and one way around it in further detail. 

Finally, for purposes of clarity of presentation, we will assume that
\begin{itemize}
\item The interaction is band-conserving.
\end{itemize}
This assumption is not necessary at all. Interactions that cause inter-band scattering (for example, a magnon-mediated spin-flip interaction) can be included in the DRP, but require keeping track of extra matrix elements. Indeed, this was done in Ref.~\cite{hutchinson2024} for the case of a bilayer material, where the presence of interlayer hopping meant that the interaction was not diagonal in the band basis. Such cases are simple to treat, but we will not consider them here in order to avoid clutter.


\section{Proof of Dimensional Reduction}
\begin{figure}
  \includegraphics[width=1.03\columnwidth]{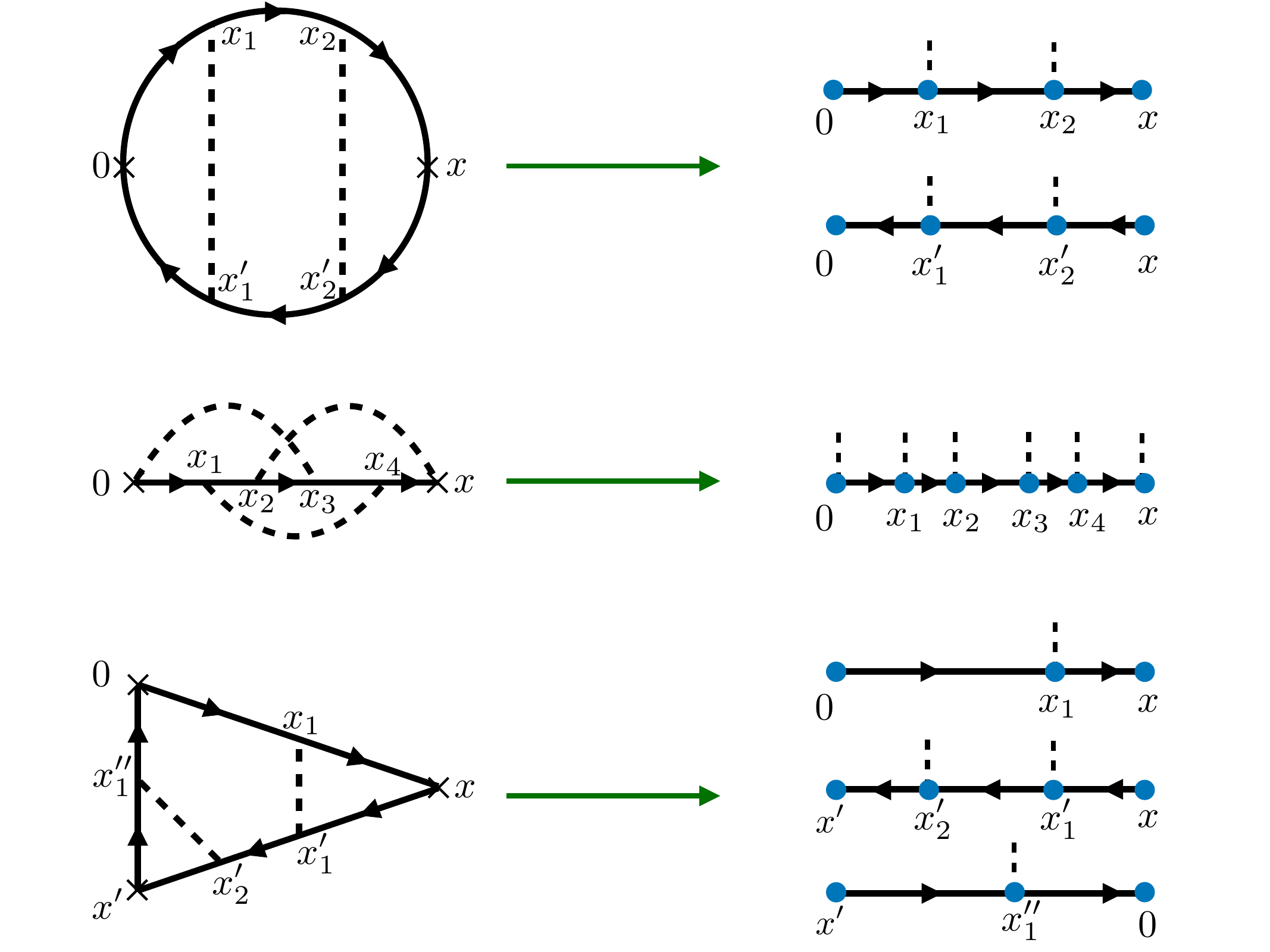}
  \caption{Examples of the decomposition of Feynman diagrams into chains for a susceptibility (top), self-energy (middle) and vertex (bottom). Crosses denote the external space-time vertices (e.g. $x\equiv (\tau,\b{r})$), solid lines denote propagators and dashed lines interactions.}
  \label{fig:chains}
\end{figure}

To facilitate the proof of the DRP, we will decompose every Feynman diagram into \emph{chains} by cutting all interaction lines and external vertices. By doing so we are (temporarily) ignoring the connectivity of the diagram. Examples of the chain decomposition are shown in Fig.~\ref{fig:chains}.

Each chain contains a product of Green's functions that connect two neighbouring vertices $g_{\sigma\nu}(\tau_{i+1}-\tau_i,\b{r}_{i+1}-\b{r}_i)$. Recall that their orientation is encoded in the $\nu$ index. Since this is always the case, it is sufficient (and convenient) to label each Green's function by it's left coordinate
\be
g_{\sigma\nu}(i)\equiv g_{\sigma\nu}(\b{r}_{i+1}-\b{r}_{i};\tau_{i+1}-\tau_{i}).
\ee

For a chain consisting of $N$ internal vertices, we denote the right-most coordinate of the chain by $x\equiv(\tau,\b{r})\equiv(\tau_{N+1},\b{r}_{N+1})$ and set the left-most coordinate to zero. In other words, we adopt the convention $g_{\sigma\nu}(N)\equiv g_{\sigma\nu}(\b{r}-\b{r}_N;\tau-\tau_N)$ and $g_{\sigma\nu}(0)\equiv g_{\sigma\nu}(\b{r}_1;\tau_1)$. Using this notation, the algebraic representation of a chain is given by
\begin{eqnarray}
C_{\sigma\nu}\left(x\right)&=&\prod_{i=1}^N\left( \int_{-\infty}^{\infty} d \tau_i \int d^D r_i g_{\sigma\nu}(i)\right)g_{\sigma\nu}(0)\nonumber\\
&&\times\{\text {interactions}\}.\label{eq:chain}
\end{eqnarray}
Due to the band-conserving assumption, matrices in the band index occur only at the external vertices. For example, if $\sigma$ is a spin index, then computing the charge susceptibility requires inserting an identity matrix at each external vertex, while the spin susceptibility requires a Pauli matrix. 

\subsection{Single chain}\label{sec:1chain}

To begin, let us focus on diagrams that consist of a single chain, i.e. self-energy diagrams. Since there is a single $\sigma$ and $\nu$ index throughout the diagram, we will drop these indices for now and restore them at the end. The goal is to first evaluate all the angular integrals in Eq.~\eqref{eq:chain}, which we will denote as
\begin{eqnarray}
\Theta(x) &\equiv& g(0) \int d \Omega_1 g(1) \int d \Omega_2 g(2) \ldots \int d \Omega_N g(N)\nonumber\\
&&\times\prod_{\{ij\}}(-1)V(|\b{r}_i-\b{r}_j|),
\end{eqnarray}
where $\Omega_i$ denotes the solid angle of the $\b{r}_i$ coordinate, and $\{ij\}$ is the set of all coordinate pairs connected by interaction lines. We have also suppressed the time coordinate of the interactions, i.e. $V(|\b{r}_i-\b{r}_j|)\equiv V(|\b{r}_i-\b{r}_j|;\tau_i-\tau_j)$, as this will not play a role in the following arguments. The $(-1)$ in front of the interaction comes from the Feynman rules.

\begin{figure}[h]
  \includegraphics[width=0.75\columnwidth]{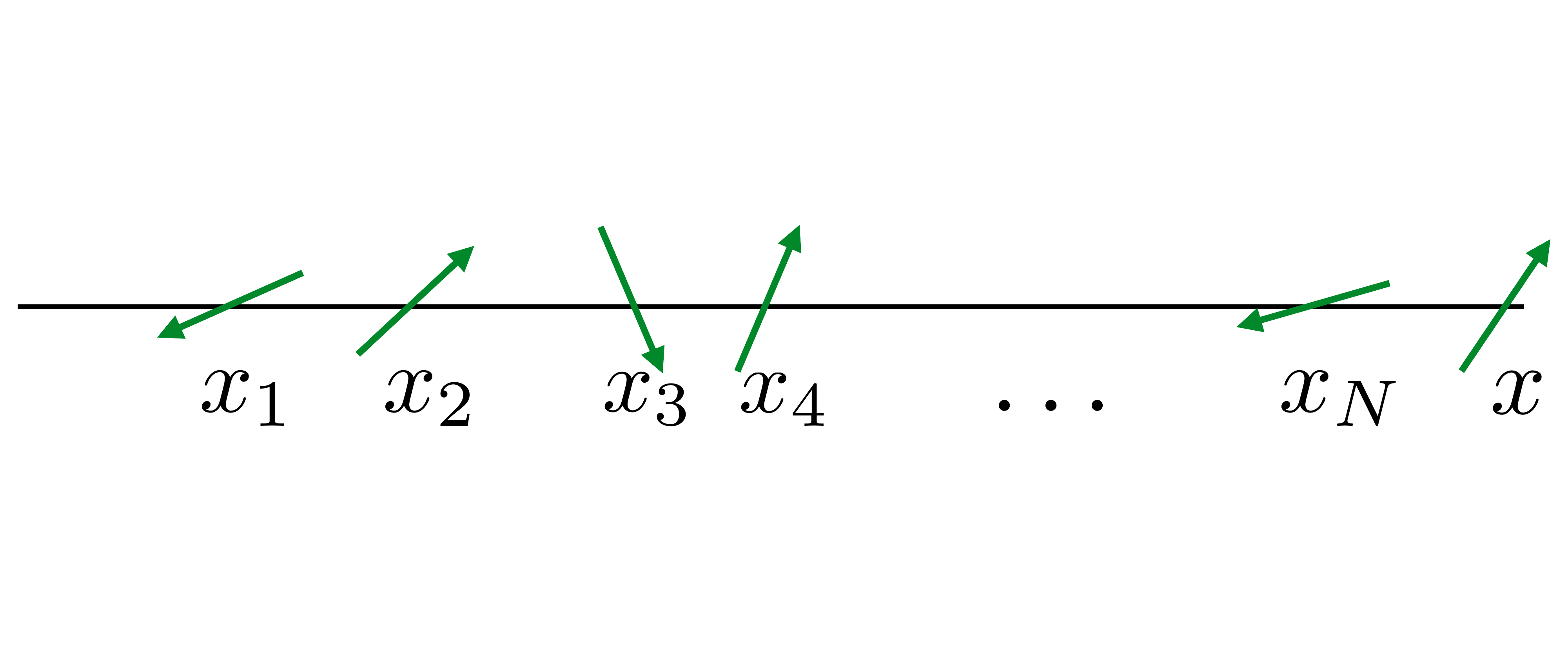}
  \caption{Rotor representation of a single chain.}
  \label{fig:rot}
\end{figure}

\begin{figure}[h]
\centering
  \includegraphics[width=1.08\columnwidth]{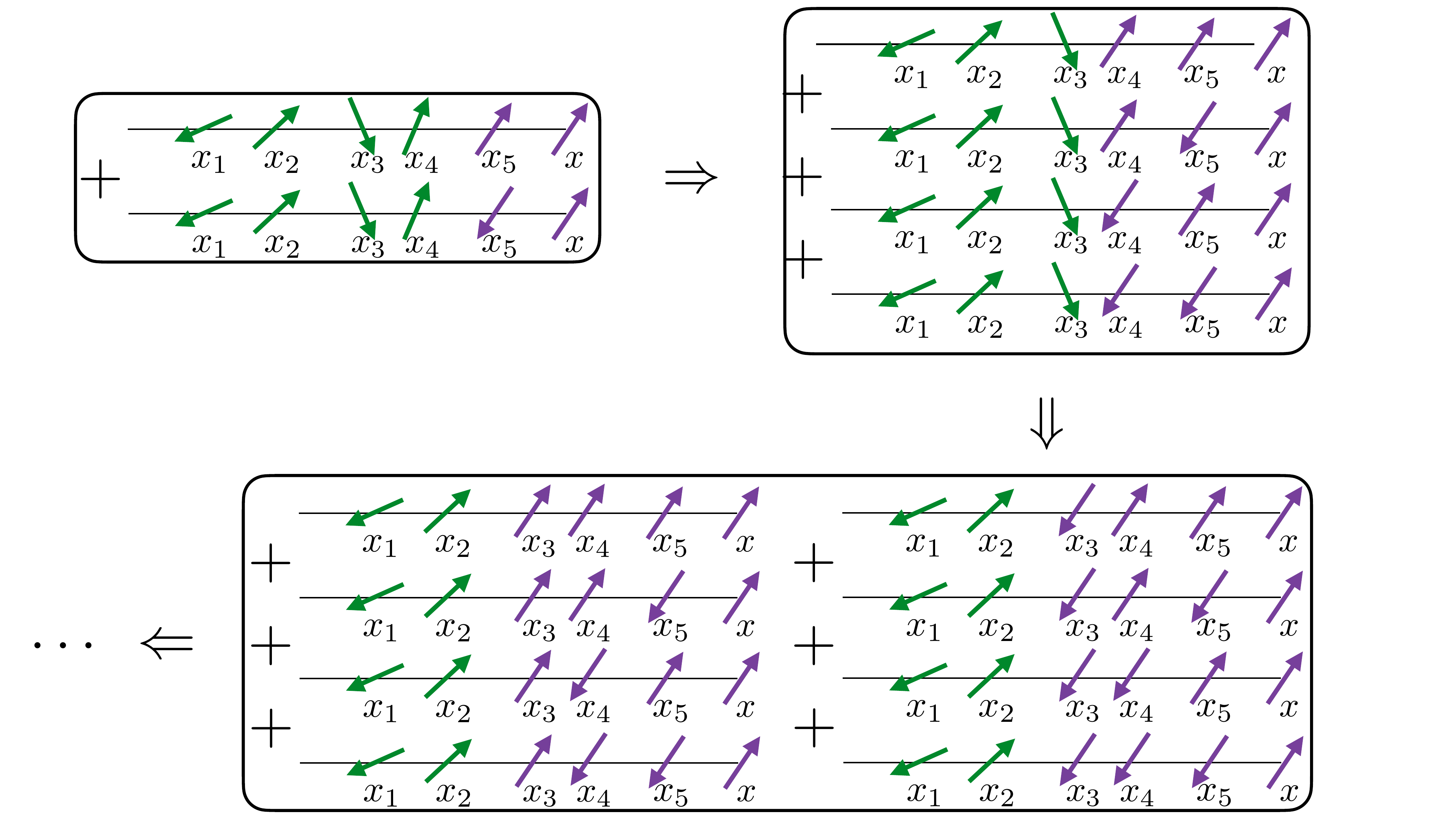}
  \caption{Example of the sequential evaluation of angular integrals for five internal vertices. Each evaluation proliferates the number of terms contributing to the chain.}
  \label{fig:rot2}
\end{figure}

To proceed, we need to separate the oscillatory and non-oscillatory parts of the integrands. For the Green's functions, the oscillatory part is simply the exponential in Eq.~\eqref{eq:Greens}. By the forward-scattering assumption, the interaction lines will not have an oscillatory component; this point is addressed in detail in Sec.~\ref{sec:back}. Thus we get 
\begin{eqnarray}
\Theta(x)&=&\sum_{s_1 \ldots s_{N+1}} e^{i s_1\left(k_F r_1-\vartheta\right)} \int d\Omega_1 e^{i s_2\left(k_F\left|\b{r}_2-\b{r}_1\right|-\vartheta\right)} \ldots \nonumber\\
&&\times\int d \Omega_N e^{i s_{N+1}\left(k_F\left|\b{r}_{N+1}-\b{r}_N\right|-\vartheta\right)}\{\text {non-osc.}\}.
\end{eqnarray}
All of these angular integrals take the form $\int d\Omega_je^{i\phi_j}$ with
\be
\phi_j\equiv s_{j+1}\left(k_F\sqrt{r_{j+1}^2+r_j^2-2r_{j+1}r_j\cos\theta_{j}}-\vartheta\right).\label{eq:phase}
\ee
Here, each $\theta_j$ is measured with respect to $\theta_{j+1}$, i.e. we align the $z$-axis of the $N$th integration with $\b{r}_{N+1}$, the $z$-axis of the $N-1$ integration with $\b{r}_N$, etc. 
The $\theta_{j}$ are represented as rotors attached to each point on the chain as in Fig.~\ref{fig:rot}. 
The integral may be evaluated via the stationary phase approximation as shown in Appendix~\ref{sec:statphase}. Eq.~\eqref{eq:phase} has two stationary points, $\theta_{j}=0,\pi$, which means 
the evaluation of the $j$th integral generates two terms: one where this rotor is aligned with its neighbour to the right, and one where it is anti-aligned. The evaluation of each angular integral thus doubles the number of terms until all rotors lie along the same axis. The first few steps in this process are shown in Fig.~\ref{fig:rot2}. 
At the end of these evaluations, we are left with a sum over all Ising-like configurations of the rotors. The relative orientation of the $j$ and $(j+1)$ rotor is encoded in an index $t_j=\pm1$. Explicitly, we can evaluate the full product of angular integrals using Eq.~\eqref{eq:Omegafinal} to get
\begin{eqnarray}
\Theta(x)&=&\sum_{s_1 \ldots s_{N+1}}e^{i s_1\left(k_F r_1-\vartheta\right)}\prod_{i=1}^N\bigg\{\sum_{t_i=\pm1}e^{is_{i+1}k_F|r_{i+1}-t_ir_i|}\nonumber\\
&\times&e^{-is_{i+1}\frac{\pi(D-1)}{4}(1-t_i)}\left(\frac{\lambda_F|r_{i+1}-t_ir_i|}{r_{i+1}r_i}\right)^{\frac{D-1}{2}}\bigg\}\nonumber\\
&\times&\{\text {non-osc.}\}'.\label{eq:thetafin0}
\end{eqnarray}
Here we have written $\{\text {non-osc.}\}'$ to remind ourselves that all the non-oscillatory functions are to be evaluated at the particular rotor configuration defined by the set of $\{t_i\}$ indices.

\begin{figure}
  \includegraphics[width=0.56\columnwidth]{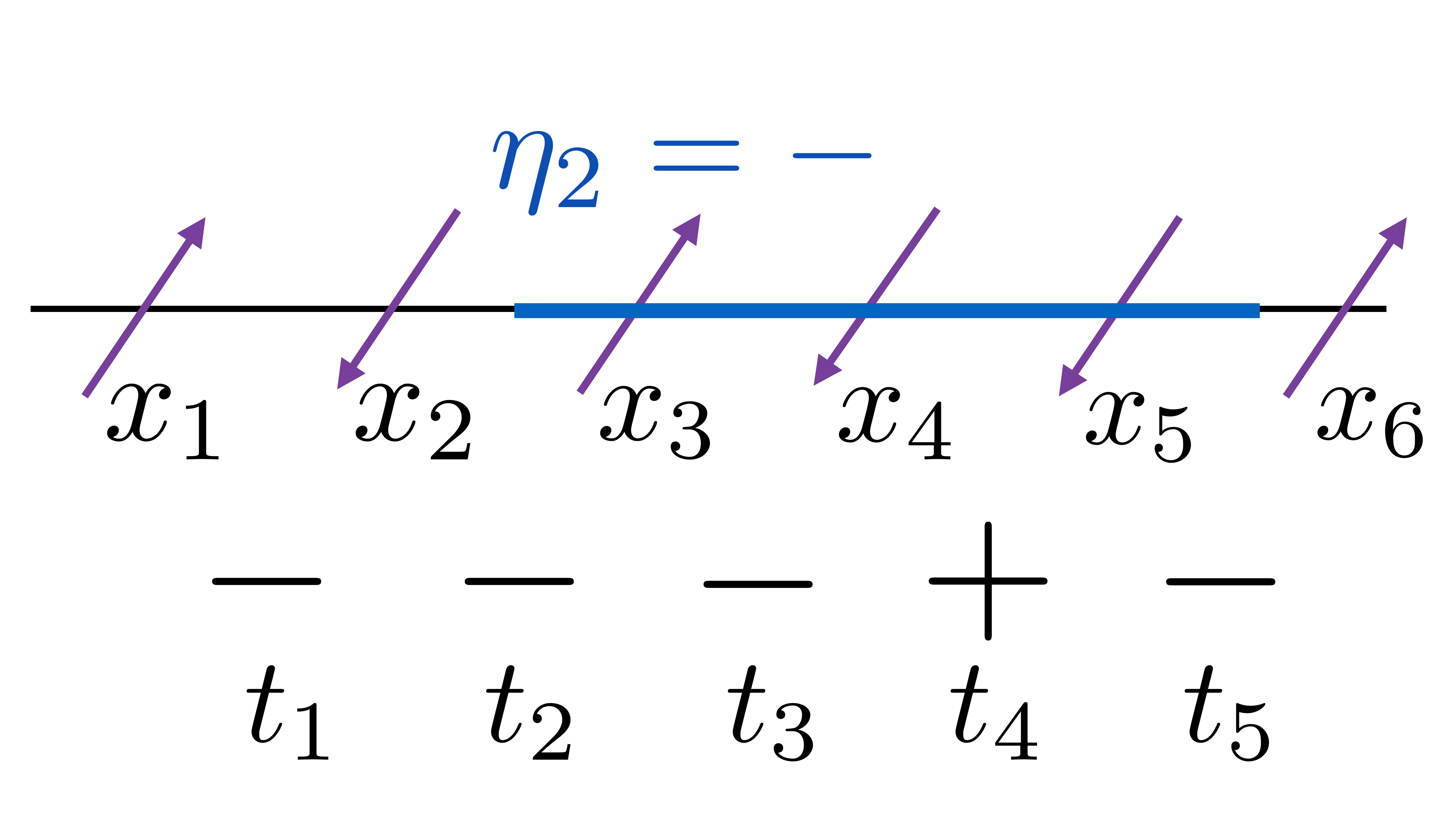}
  \caption{An example of a six rotor Ising configuration with $t_i$ values labelled. The Ising index $\eta_2$ is indicated in blue.}
  \label{fig:rot3}
\end{figure}

A moment's reflection reveals that we have not indexed the different rotor configurations in the most natural way since knowing the orientation of rotor $i$ requires not only knowing $t_i$, but also the orientation of rotor $i+1$. It is much more convenient to measure the absolute orientation of each rotor with respect the final ($N+1$) rotor in the chain. To that end, we define the ``Ising index'' $\eta_i$ as
\be
\eta_i\equiv\prod_{j=i}^Nt_j.
\ee 
An example of the relation between the $\eta$'s and $t$'s is shown in Fig.~\ref{fig:rot3}.

In terms of the Ising indices, Eq.~\eqref{eq:thetafin0} becomes
\begin{eqnarray}
\Theta(x)&=&\sum_{s_1 \ldots s_{N+1}}e^{i s_1\left(k_F r_1-\vartheta\right)}\nonumber\\
&\times&\prod_{i=1}^N\bigg\{\sum_{\eta_i=\pm1}e^{is_{i+1}\eta_{i+1}k_F|\eta_{i+1}r_{i+1}-\eta_ir_i|}\nonumber\\
&\times&e^{-is_{i+1}\frac{\pi(D-1)}{4}(1-\eta_i\eta_{i+1})}\nonumber\\
&\times&\left(\frac{\lambda_F|\eta_{i+1}r_{i+1}-\eta_ir_i|}{r_{i+1}r_i}\right)^{\frac{D-1}{2}}\bigg\}\{\text {non-osc.}\}'.\label{eq:thetafin}
\end{eqnarray}

To proceed further we must recognize that this expression will be integrated over all internal radial coordinates $r_i$ once it is inserted in the expression for the chain Eq.~\eqref{eq:chain}. For most configurations (i.e. most values of $s_i$ and $\eta_i$), the phase will depend on $r_i$ so that the oscillating integrand will nearly cancel upon performing the radial integrations. However, there are some configurations where the phase is independent of all internal radial coordinates, and these will provide the leading contribution. In the product over $i$ in Eq.~\eqref{eq:thetafin}, any given internal coordinate $r_j$ appears in two consecutive phase factors (coming from the two Green's functions connected to that vertex). It's coefficient vanishes if and only if
\be
s_j\eta_j\sgn(\eta_jr_j-\eta_{j-1}r_{j-1})=s_{j+1}\sgn(\eta_{j+1}r_{j+1}-\eta_jr_j)\eta_j.
\ee
This is, in fact, a recursion relation for the chiral indices. If we apply this relation $j$ times, we can write all chiral indices in terms of the first one $s\equiv s_1$,
\be
s_{j+1}=\sgn(\eta_{j+1}r_{j+1}-\eta_jr_j)\eta_1s.\label{eq:rec}
\ee

Thus the leading contribution to the radial integrals comes from
\begin{eqnarray}
\Theta(x)&\approx&\sum_{\eta_1,\ldots,\eta_N}\sum_{s}e^{is\rho_N[k_Fr-\frac{\pi}{4}(D-1)]}\nonumber\\
&\times&\prod_{i=1}^N\left(\frac{\lambda_F|\eta_{i+1}r_{i+1}-\eta_ir_i|}{r_{i+1}r_i}\right)^{\frac{D-1}{2}}\{\text {non-osc.}\}',\nonumber\\
\label{eq:thetafin2}
\end{eqnarray}
where we have defined
\begin{eqnarray}
\rho_N&\equiv&1+\sum_{i=1}^N\sgn(\eta_{i+1}r_{i+1}-\eta_ir_i)\eta_1(1-\eta_i\eta_{i+1})\label{eq:rho}.\nonumber\\
\end{eqnarray}

The oscillatory parts of Eq.~\eqref{eq:thetafin2} have all been taken care of except for the unusual-looking phase $\rho_N$ (Eq.~\eqref{eq:rho}). But in fact, we can prove by induction that
\be
\rho_N=\eta_1.
\ee
\begin{proof}
First we establish the mathematical identity $\rho_N=\eta_1\eta_{N+1}$, valid regardless of whether or not $N$ refers to the length of the chain. For $N=1$, we have
\begin{eqnarray}
\rho_1&=&1+\sgn(\eta_2r_2-\eta_1r_1)\eta_1(1-\eta_1\eta_2)\\
&=&\begin{cases}
1-2\sgn(r_2+r_1) & \eta_2=-\eta_1\\
1 & \eta_2=\eta_1
\end{cases}\\
&=&\eta_1\eta_2,
\end{eqnarray}
since $r_1$ and $r_2$ are radial coordinates and therefore positive. Now assume the claim holds for some $N-1$, then for $N$ we have
\begin{eqnarray}
\rho_N&=&\rho_{N-1}+\sgn(\eta_{N+1}r_{N+1}-\eta_{N}r_{N})\eta_1(1-\eta_{N}\eta_{N+1})\nonumber\\
\\
&=&\eta_1\eta_N+\begin{cases}
2\eta_{N+1}\eta_1& \eta_{N}=-\eta_{N+1}\\
0 &\eta_{N}=\eta_{N+1}
\end{cases}\\
&=&\eta_1\eta_{N+1}.
\end{eqnarray}
The last rotor in the array is always aligned with itself so that $\eta_{N+1}=1$, which completes the proof.
\end{proof}

Let us now be specific about the non-oscillatory factors in Eq.~\eqref{eq:thetafin2}. These come from the prefactors and denominator of each Green's function in Eq.~\eqref{eq:Greens} as well as the interactions, all evaluated at a given Ising configuration. Since the interaction lines may connect any two coordinates, $\b{r}_i$ and $\b{r}_j$, we need to know their relative orientation. 
After applying the stationary phase approximation, each $\b{r}_i$ becomes $\eta_ir_i$ and interaction lines take the form $V(|\eta_ir_i-\eta_jr_j|)$ so that
\begin{eqnarray}
\{\text {non-osc.}\}'&=&\frac{g^{({\rm 1D})}(0)}{(\lambda_F r_1)^{\frac{D-1}{2}}}\prod_{i=1}^N\frac{g^{({\rm 1D})}(i)}{(\lambda_F |\eta_{i+1}r_{i+1}-\eta_ir_i|)^{\frac{D-1}{2}}}\nonumber\\
&&\times\prod_{\{jk\}}(-1)V(|\eta_jr_j-\eta_kr_k|).\label{eq:nonosc}
\end{eqnarray}
Here 
we have suggestively defined
\begin{eqnarray}
g^{({\rm 1D})}(i)&\equiv&\frac{1}{2\pi}\frac{1}{is_{i+1}|\eta_{i+1}r_{i+1}-\eta_ir_{i}|-v_F(\tau_{i+1}-\tau_i)}.\nonumber\\
\end{eqnarray}
We can use the recursion relation~\eqref{eq:rec} to deal with the $s_{i+1}$ factor in the denominator. Furthermore, noting that $\eta_1=\pm1$ and the chiral index $s$ only appears in the combination $\eta_1 s$, we may redefine the chiral index as $s\rightarrow \eta_1 s$ to get
\begin{eqnarray}
g^{({\rm 1D})}(i)=\frac{1}{2\pi}\frac{1}{is(\eta_{i+1}r_{i+1}-\eta_ir_{i})-v_F(\tau_{i+1}-\tau_i)}.\label{eq:g1d}\nonumber\\
\end{eqnarray} 

We are now ready to write down the radial integrals explicitly. Note that the denominator $(\lambda_F |\eta_{i+1}r_{i+1}-\eta_ir_i|)^{\frac{D-1}{2}}$ in Eq.~\eqref{eq:nonosc} precisely cancels the corresponding factor in Eq.~\eqref{eq:thetafin2} produced by evaluating the angular integrals. 
So the full expression for the chain (Eq.~\eqref{eq:chain}) becomes
\begin{eqnarray}
C\left(x\right)&=&\sum_{\eta_1\ldots\eta_N}\sum_se^{is[k_Fr-\frac{\pi}{4}(D-1)]}\nonumber\\
&\times&\prod_{i=1}^N\left(\int_{-\infty}^{\infty} d \tau_i \int_0^\infty dr_i r_i^{D-1}\frac{g^{({\rm 1D})}(i)}{(r_{i+1}r_i)^{\frac{D-1}{2}}}\right)\nonumber\\
&\times& \frac{g^{({\rm 1D})}(0)}{(\lambda_F r_1)^{\frac{D-1}{2}}}\prod_{\{jk\}}(-1)V(|\eta_jr_j-\eta_kr_k|).\label{eq:chainfin}
\end{eqnarray}
Furthermore, all of the radial Jacobian factors precisely cancel the $r_i$ factors in the denominator with the exception of $r_{N+1}=r$. This simplifies the chain expression greatly, but there is one more critical realization that will allow us to make the connection to 1D: the entire summand in Eq.~\eqref{eq:chainfin} is invariant under the simultaneous transformation $\eta_i\rightarrow-\eta_i$, $r_i\rightarrow-r_i$. This means that instead of summing over all possible Ising configurations $\eta_i=\pm1$, we can choose a single representative configuration (say the fully parallel case $\eta_i=1$ for all $i$) and instead extend every radial integral to $-\infty$! Restoring auxiliary indices and the time argument of the interactions gives the final expression for the chain
\begin{eqnarray}
C_{\sigma\nu}\left(x\right)&=&\sum_s\frac{e^{is[k^\sigma_Fr-\frac{\pi}{4}(D-1)]}}{(\lambda_F^\sigma r)^{\frac{D-1}{2}}}\nonumber\\
&\times&\prod_{i=1}^N\left(\int_{-\infty}^{\infty} d \tau_i \int_{-\infty}^\infty dr_ig^{({\rm 1D})}_{\sigma\nu s}(i)\right)g^{({\rm 1D})}_{\sigma\nu s}(0)\nonumber\\
&\times&\prod_{\{jk\}}(-1)V(|r_j-r_k|;\tau_j-\tau_k),\label{eq:chainfin2}
\end{eqnarray}
where 
\be
g^{({\rm 1D})}_{\sigma\nu s}(i)=\frac{1}{2\pi}\frac{1}{i s(r_{i+1}-r_{i})-v_F^{\sigma\nu}(\tau_{i+1}-\tau_i)}.\label{eq:g1dfin}
\ee
Now we can see the reason for the name ``chiral index,'' since Eq.~\eqref{eq:g1dfin} is none other than the one-dimensional Green's function in the Tomonaga-Luttinger model for left-movers ($s=-1$) and right-movers ($s=+1$)~\cite{tomonaga1950}.

With the exception of the prefactor on the the first line, Eq.~\eqref{eq:chainfin2} is exactly the algebraic expression for the equivalent chain diagram in one dimension, thus completing the dimensional-reduction proof for one chain.

\subsection{Two chains}\label{sec:2chainz}
\begin{figure}[h]
  \includegraphics[width=0.86\columnwidth]{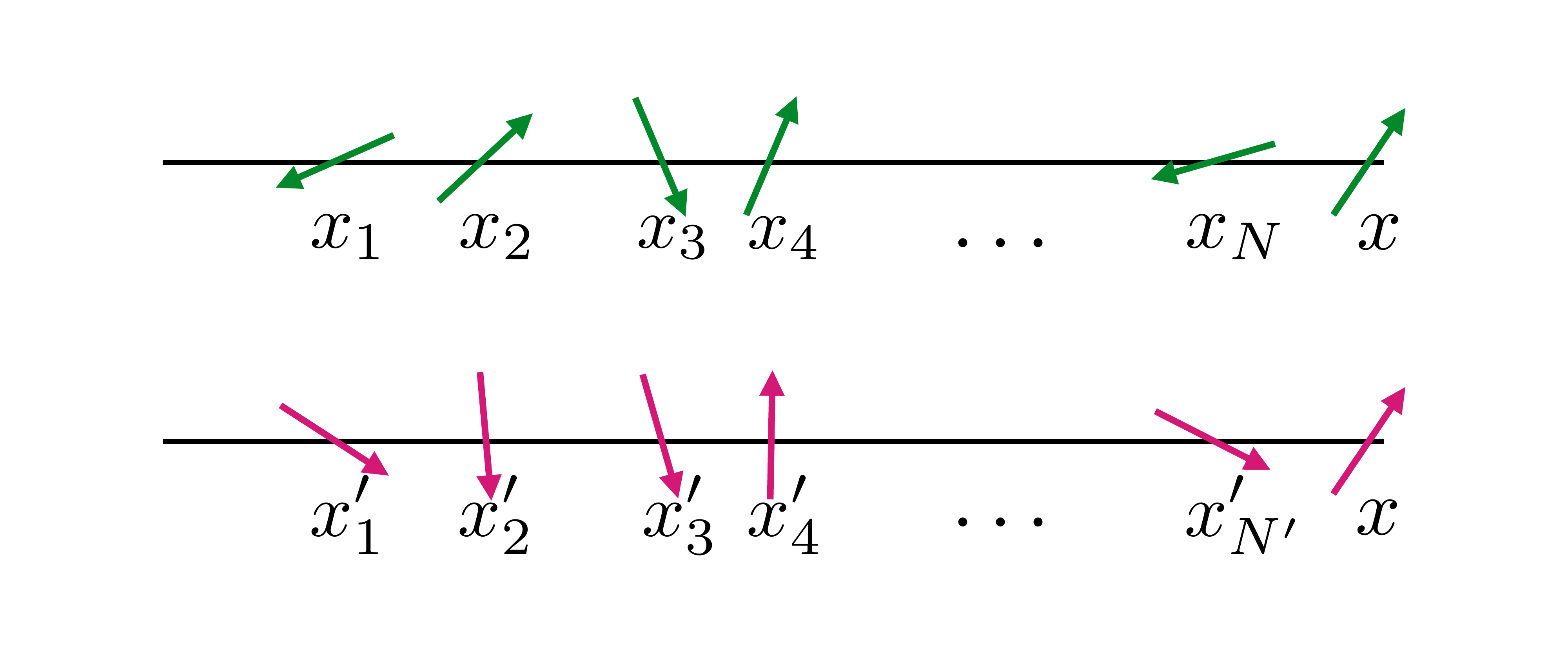}
  \caption{Rotor representation of two chains.}
  \label{fig:rot4}
\end{figure}

What if our diagram of interest is composed of two chains such as the susceptibility shown in the top of Fig.~\ref{fig:chains}? The second chain will have a number of vertices $N'$ which may be different than the first chain. If they are not connected by any interaction lines, then clearly we can apply the same procedure as in section~\ref{sec:1chain}. We concatenate the two independent chains and the result is once again given by the appropriate linear combination of one-dimensional  diagrams made of left and right-movers. Do interactions that connect the two chains spoil this? No. The reason is that we consider only forward-scattering interactions. Thus there are no angular integrations that involve the relative orientation of two rotors on different chains. So the stationary-phase approximation proceeds unimpeded. Moreover, because the endpoint of both chains is the same (see Fig.~\ref{fig:rot4}), all Ising configurations occur along a single universal axis. Thus an interaction connecting the two chains $V(|\b{r}_i-\b{r}'_j|)$ becomes $V(|\eta_ir_i-\eta_j'r'_j|)$ after evaluating the angular integrals. Both sets of Ising indices $\eta_i$ and $\eta_i'$ are summed over, which, as we saw before, is equivalent to extending all the radial integrals to the entire real line. 

This means chains combine in the obvious way. Explicitly, if we extend our vertex indices via
\be
r_i\equiv\begin{cases}
r_i & 1\leq i\leq N\\
r_i' & N< i\leq N+N'
\end{cases},
\ee
then a generic susceptibility diagram can be written as
\begin{eqnarray}
C_{\sigma\nu}^{\sigma'\nu'}&=&\sum_{ss'}\frac{e^{is[k^\sigma_Fr-\frac{\pi}{4}(D-1)]}e^{is'[k^{\sigma'}_Fr-\frac{\pi}{4}(D-1)]}}{\left(\sqrt{\lambda_F^\sigma\lambda_F^{\sigma'}}r\right)^{D-1}}\nonumber\\
&\times&\prod_{i=1}^{N}\left(\int_{-\infty}^{\infty} d \tau_i \int_{-\infty}^\infty dr_ig^{({\rm 1D})}_{\sigma\nu s}(i)\right)g^{({\rm 1D})}_{\sigma\nu s}(0)\nonumber\\
&\times&\prod_{i=N+1}^{N+N'}\left(\int_{-\infty}^{\infty} d \tau_i \int_{-\infty}^\infty dr_ig^{({\rm 1D})}_{\sigma'\nu' s'}(i)\right)g^{({\rm 1D})}_{\sigma'\nu' s'}(0)\nonumber\\
&\times&\prod_{\{jk\}}(-1)V(|r_j-r_k|; \tau_j-\tau_k).\label{eq:susceptfin}
\end{eqnarray}

\begin{figure*}[t]
  \includegraphics[width=1.88\columnwidth]{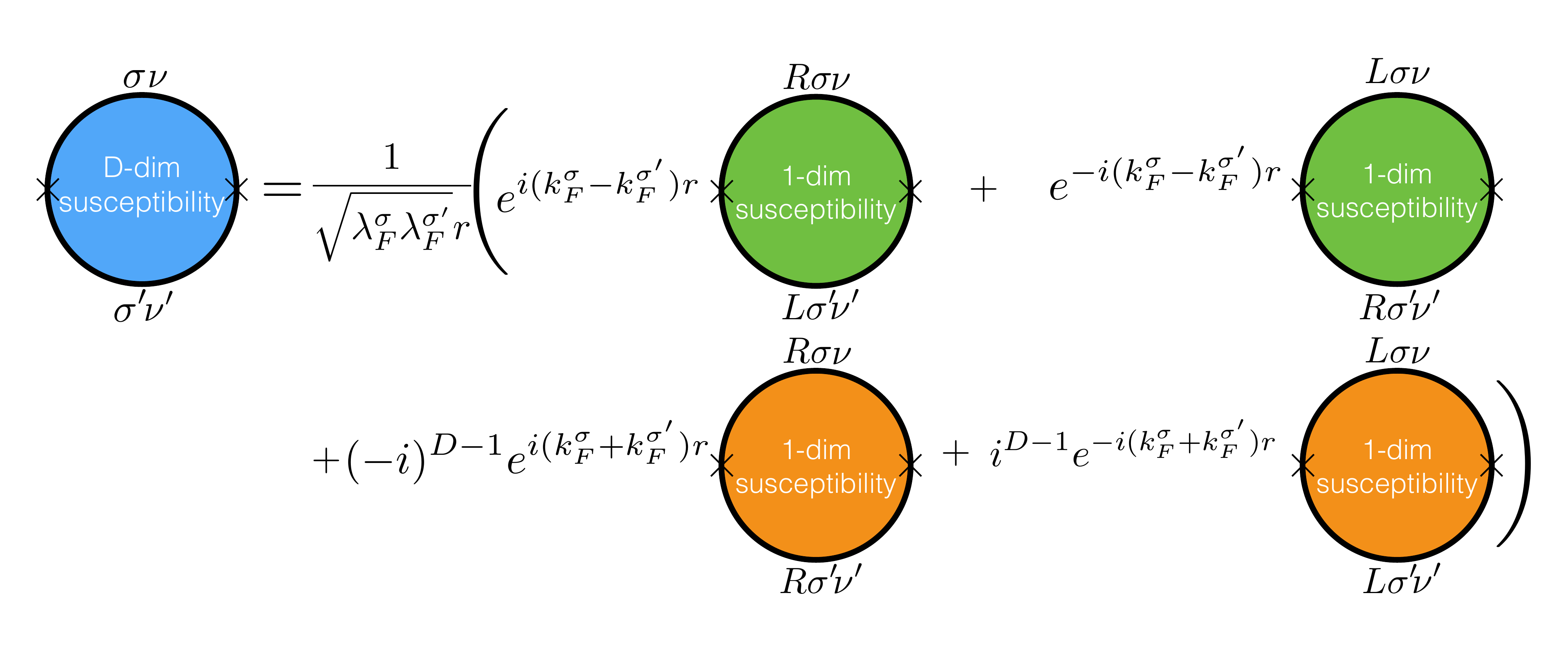}
  \caption{Exact statement of dimensional reduction for susceptibility diagrams. The $D$-dimensional irreducible susceptibility on the left (with solid lines indicating products of $g_{\sigma\nu}$ Green's functions on top, and products of $g_{\sigma'\nu'}$ on the bottom) is equivalent to the linear combination of the corresponding one-dimensional susceptibilities on the right (with solid lines indicating products of left (L)- and right (R)-moving 1D Green's functions). The green (Landau) terms have prefactors that are roughly constant (for small Fermi-surface splitting) and contribute to the $q=0$ physics. The orange (Kohn) terms have prefactors that oscillate close to $q=2k_F$ and contribute to instabilities there.}
  \label{fig:dimred}
\end{figure*}

Diagramatically, this equation is illustrated in Fig.~\ref{fig:dimred}. This is the essential DRP result. It says that a $D$-dimensional irreducible susceptibility diagram can be expressed as  the sum of four one-dimensional diagrams (coming from the sum over $s,s'=\pm1$). Two of these oscillate slowly as a function of $r$, while two of them oscillate near $2k_F$. 

\section{Backscattering}\label{sec:back}
It is important to understand the significance of the forward-scattering assumption. In general, an interaction without this restriction has an oscillatory part given by its Fourier transform
\be
V(\b{r}_i-\b{r}_j)=\int\frac{d^dk}{(2\pi)^d}e^{i\b{k}\cdot(\b{r}_i-\b{r}_j)}V(\b{k}).
\ee
This additional phase factor will complicate the stationary-phase approximation. The stationary points are still captured by some  Ising configuration ($\b{r}_i$ and $\b{r}_j$ parallel or anti-parallel), but the quadratic fluctuation terms will no longer integrate to the simple values given in Eq.~\eqref{eq:Omegafinal} since the angle $\theta_i$ appears both in the Green's function phase factor and the interaction phase factor. Nonetheless, one can in principle compute these integrals. The result in Eq.~\eqref{eq:thetafin} will be more complicated, but the oscillatory part will just gain an extra exponential of the form $e^{ik|r_i\pm r_j|}$. Which values of $k$ are relevant? We once again recognize that the leading contribution comes from the phase factors that are independent of $r_i$. Each of the two Green's function attached to a vertex contributes a factor of $e^{\pm ik_Fr_i}$. To compensate this, the interaction must either contribute no phase factor (forward scattering) or a factor of $e^{\pm i2k_Fr_i}$ (backscattering). Thus by requiring $V(2k_F)\ll V(0)$, we can ignore this pesky case.

\begin{figure}[h]
  \includegraphics[width=0.87\columnwidth]{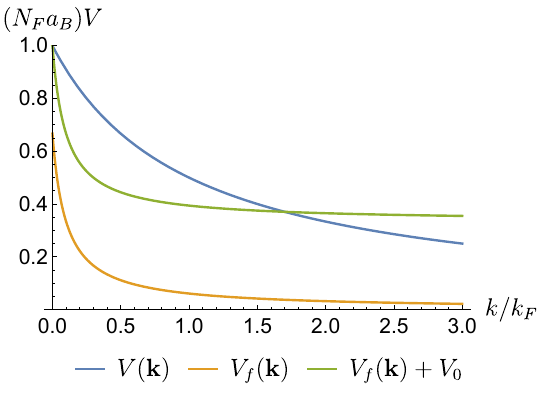}
  \caption{An example of approximating a Thomas-Fermi interaction with $k_FR_s=1$ (blue) by a forward-scattering part with $k_FR_s'=10$ (orange) and a constant $V_0=(N_Fa_B)^{-1}/(2k_F+\frac{1}{R_s})$. The approximation is shown in green.} 
  \label{fig:vapprox}
\end{figure}

This assumption is necessary for the dimensional reduction, but is there a way to restore the backscattering physics in a simple way? In fact, there is a way, and this restoration has some profound consequences. The idea is to take the full interaction $V(\b{k})$ and decompose it into a forward-scattering piece and a constant tail, i.e.
\be
V(\b{k})\approx V_f(\b{k})+V_0,
\ee  
where $V_f(2k_F)\ll V_f(0)$. To be concrete, suppose we consider a Thomas-Fermi interaction in 2D
\be
V(\b{k})=\frac{(N_Fa_B)^{-1}}{k+R_s^{-1}},
\ee
where $N_F=\frac{m_*}{2\pi\hbar^2}$ is the density of states for a spin-degenerate band, $a_B=\frac{\epsilon}{m_*e^2}$ is the effective Bohr radius and $R_s$ is the screening length. If $k_FR_s\gg1$, then $V(0)\gg V(2k_F)$, the interaction is of forward-scattering type and dimensional reduction proceeds unimpeded. This is indeed the case for some materials, but it is also possible that $k_FR_s\sim1$, in which case it is not clear if the dimensionally reduced result will give the correct analytic behaviour of correlation functions. In this case, we define $V_0\equiv V(2k_F)$, and
\be
V_f(\b{k})\equiv\frac{R_s}{R_s'}\left(\frac{2k_FR_s}{2k_FR_s+1}\right)\frac{(N_Fa_B)^{-1}}{k+1/R_s'},
\ee
where the new screening length $R_s'$ satisfies the forward-scattering condition $k_FR_s'\gg1$. The prefactor is chosen such that $V_f(0)+V_0=V(0)$. 
This approximation is shown in Fig.~\ref{fig:vapprox}. The goal here is not to accurately approximate the interaction we started with, but rather to ensure that we maintain both the long-range and short-range physics in our diagrams. 

\begin{figure}[h]
  \includegraphics[width=0.87\columnwidth]{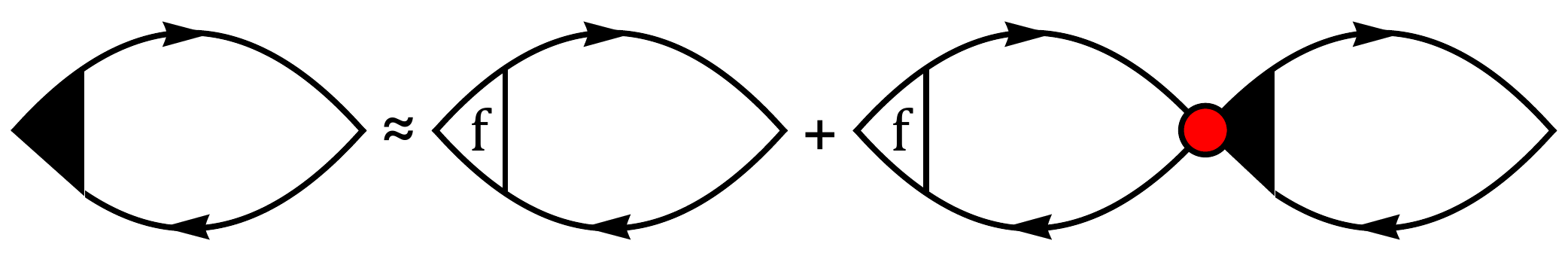}
  \caption{The full susceptibility in Nambu space $\chi$ represented by the bubble with the black triangle, is approximated by the Cooper ladder where each vertex is dressed by the finite-range interaction $V_f$. This dressing results in the forward-scattering susceptibility $\chi_f$ (bubble with the white triangle denoted by ``f''). These bubbles are linked together via the contact interaction $V_0$ denoted by the red circle.}
  \label{fig:ladder}
\end{figure}

Now we can treat the contact part of the interaction by standard means. For example, suppose we are interested in the Cooper instability of the pair-susceptibility diagram. $V_0$ is then nothing but the usual Bardeen-Cooper-Schrieffer interaction, for which it is standard to consider the ladder series for the susceptibility. This series effectively links together a sequence of bare susceptibility diagrams. Each of these can be dressed by the forward-scattering part $V_f(\b{k})$ and evaluated using the DRP to obtain $\chi_f(\b{q},\omega)$. This process, illustrated in Fig.~\ref{fig:ladder}, gives an integral equation for the pair susceptibility, whose solution is
\be
\chi(\b{q},\omega)=\frac{\chi_f(\b{q},\omega)}{1+V_0\chi_f(\b{q},\omega)}.
\ee

This approach was used in Ref.~\cite{miserev2025} to study the superconducting instability originating from optical-phonon interactions in layered compounds. Note that the backscattering term makes it easier for the Fermi surface to destabilize, since it is no longer necessary that $\chi_f$ diverges at some $\b{q}$, but rather just that $\chi_f(\b{q},\omega)=-1/V_0$.
\section{Simple examples}
\subsection{Zeroth-order static susceptibiltiy}
The simplest two-chain example we can consider is the bare static susceptibility $\chi_0(\b{q},\omega=0)$. Even without interactions, this is a non-trivial calculation but can be found in many textbooks. We show here that the DRP captures the correct non-analyticities of this function in every dimension.

The exact results are given by~\cite{mihaila2011}
\begin{eqnarray}
\chi_0^{\rm 1D}(q,0)&=&-\frac{2k_F}{\pi v_F}\frac{1}{q}\ln\left|\frac{1+\frac{q}{2k_F}}{1-\frac{q}{2k_F}}\right|\label{eq:1Dref}\\
\chi_0^{\rm 2D}(q,0)&=&-\frac{2}{v_F\lambda_F}\left[1-\Theta(q-2k_F)\sqrt{\frac{q}{k_F}-2}\right]\label{eq:2Dref}\\
\chi_0^{\rm 3D}(q,0)&=&-\frac{k_F}{\pi v_F\lambda_F}\left[1+\frac{1-(\frac{q}{2k_F})^2}{q/k_F}\ln\left|\frac{1+\frac{q}{2k_F}}{1-\frac{q}{2k_F}}\right|\right],\nonumber\\
\label{eq:3Dref}
\end{eqnarray}
where $\Theta(x)$ is the Heaviside step function.

Using DRP, we can read off the real-space result directly from Eq.~\eqref{eq:susceptfin}~\footnote{Note that $\chi_0=-\Pi_0$, where $\Pi_0$ is the polarization. $\Pi_0$ is given by the bubble diagram itself, which has a minus sign from the fermion loop. The DRP result comes from gluing two chains together and does not account for this minus sign, hence it gives $\chi_0$ rather than $\Pi_0$.}
\begin{eqnarray}
\chi_0(\b{r},\tau)&=&2\sum_{ss'}\frac{e^{is[k_Fr-\frac{\pi}{4}(D-1)]}e^{is'[k_Fr-\frac{\pi}{4}(D-1)]}}{\left(\lambda_Fr\right)^{D-1}}\nonumber\\
&\times&g^{({\rm 1D})}_{e s}(0)g^{({\rm 1D})}_{h s'}(0),
\end{eqnarray}
where we traced over the spin indices to get the prefactor of 2 and the $e/h$ subscript denotes electron/hole propagators ($\nu=\pm1$). 

Eq.~\eqref{eq:g1dfin} gives
\be
g^{({\rm 1D})}_{(e/h) s}(0)=\frac{1}{2\pi}\frac{1}{i sr\mp v_F\tau}.
\ee
All that remains is to compute the Fourier transform. The time integral of the Green's functions is
\begin{eqnarray}
T(r)&\equiv&\int_{-\infty}^\infty d\tau g^{({\rm 1D})}_{e s}(0)g^{({\rm 1D})}_{h s'}(0)\nonumber\\
&=&\int_{-\infty}^\infty \frac{d\tau}{(2\pi)^2} \frac{1}{(isr-v_F\tau)(is'r+v_F\tau)}.
\end{eqnarray}
The only non-zero contributions occur if the poles are in opposite half-planes, i.e. $s=s'$, for which the residue theorem gives
\be
T(r)=-\frac{1}{4\pi v_F r}\delta_{ss'},
\ee 
so that
\be
\chi_0(\b{r},\omega=0)=-\frac{\cos[2k_Fr-\frac{\pi}{2}(D-1)]}{\pi v_F\lambda_F^{D-1}r^D}.
\ee
In momentum space,
\be
\chi_0(\b{q},0)=\frac{-1}{\pi v_F\lambda_F^{D-1}}\int d^Dr\frac{\cos[2k_Fr-\frac{\pi}{2}(D-1)]}{r^D}e^{i\b{q}\cdot\b{r}}.
\ee 
The angular integrals can be evaluated using Eq.~\eqref{eq:Ftfinal} so that 
\begin{eqnarray}
\chi_0(q,0)&=&\frac{-2}{v_F\lambda_F^{D-1}}\left(\frac{2\pi}{q}\right)^{D/2-1}\int_0^\infty \frac{dr}{r^{D/2}}J_{D/2-1}(qr)\nonumber\\
&&\times\cos[2k_Fr-\frac{\pi}{2}(D-1)].
\end{eqnarray}
The remaining integral can be evaluated analytically in one, two and three dimensions. For one and three dimensions, it must be regularized by a short-distance cutoff $\Lambda$. This is an artifact of the asymptotic approximation, which fails to capture the short-distance physics.
\begin{itemize}
\item \emph{1D:}
\begin{eqnarray}
\chi_0(q,0)&=&\frac{-2}{v_F}\sqrt{\frac{q}{2\pi}}\int_\Lambda^\infty \frac{dr}{r^{1/2}}J_{-1/2}(qr)\cos(2k_Fr)\nonumber\\
\\
&=&\frac{-1}{\pi v_F}\bigg[\mathcal{C}[(2k_F+q)r]+\mathcal{C}[(2k_F-q)r]\bigg]_\Lambda^\infty,\nonumber\\
\label{eq:chi1d}
\end{eqnarray}
where $\mathcal{C}(x)$ denotes the cosine integral. Non-analyticies of all the bare bubbles occur at $q=2k_F$. Expanding \eqref{eq:chi1d} around this point gives
\begin{eqnarray}
\chi_0(q,0)&\approx&\frac{1}{\pi v_F}\left(\mathcal{C}[(4k_F)\Lambda]+\gamma+\ln|(2k_F-q)\Lambda|\right)\nonumber\\
&=&\frac{1}{\pi v_F}\ln\left|1-\frac{q}{2k_F}\right|+{\rm const.}.\label{eq:chi1dfinal}
\end{eqnarray}
The DRP is unable to produce the correct constant offsets, but the important information contained in the singularity of~\eqref{eq:chi1dfinal} agrees exactly with the full result~\eqref{eq:1Dref}. 

\item \emph{2D:}
\begin{eqnarray}
\chi_0(q,0)&=&\frac{-2}{v_F\lambda_F}\int_0^\infty \frac{dr}{r}J_{0}(qr)\sin(2k_Fr)\\
&=&\frac{-2}{v_F\lambda_F}\begin{cases}
\frac{\pi}{2} & q<2k_F\\
\arcsin\left(\frac{2k_F}{q}\right) & q>2k_F
\end{cases}
\end{eqnarray}
Expanding around $q=2k_F$ gives
\begin{eqnarray}
\chi_0(q,0)&\approx&\frac{-2}{v_F\lambda_F}\left[\frac{\pi}{2}-\Theta(q-2k_F)\sqrt{\frac{q}{k_F}-2}\right],
\end{eqnarray}
which also agrees with the exact result~\eqref{eq:2Dref} up to a constant shift.

\item\emph{3D:}
\begin{eqnarray}
\chi_0(q,0)&=&\frac{2}{v_F\lambda_F^{2}}\sqrt{\frac{2\pi}{q}}\int_\Lambda^\infty \frac{dr}{r^{3/2}}J_{1/2}(qr)\cos(2k_Fr)\nonumber\\
\\
&=&\frac{-2}{v_F\lambda_F^{2}}\frac{1}{q}\bigg[(q-2k_F)\mathcal{C}[(2k_F-q)r]\nonumber\\
&&+(q+2k_F)\mathcal{C}[(2k_F+q)r]\nonumber\\
&&+\frac{\sin[(2k_F-q)r]}{r}-\frac{\sin[(2k_F+q)r]}{r}\bigg]_{\Lambda}^\infty.\nonumber\\
\end{eqnarray}
Near $q=2k_F$ this gives (up to a constant)
\be
\chi_0(q,0)\approx\frac{-1}{2\pi v_F\lambda_F}(q-2k_F)\ln\left|1-\frac{q}{2k_F}\right|,
\ee
again in agreement with the exact result~\eqref{eq:3Dref}.
\end{itemize} 

\subsection{Fock self-energy}
The previous examples did not include any interaction lines. To test the DRP on a simple diagram with an interaction line, we consider the first-order Fock self-energy $\Sigma^{({\rm F})}$ with a Coulomb interaction $V(r)=e^2/(\epsilon r)$, which automatically satisfies the forward-scattering condition $V(0)\gg V(2k_F)$. Here $e$ is the elementary charge and $\epsilon$ is the dielectric constant.
 
In three dimensions, for a spherical Fermi surface, the exact result is given by~\cite{mahan2000}
\begin{equation}
\Sigma^{({\rm F})} (\b{k},\omega)= -\frac{e^2 k_F}{\pi\epsilon}\left[1+\frac{1}{2}\left(\frac{k_F}{k}-\frac{k}{k_F}\right)\ln\left|\frac{k+k_F}{k-k_F}\right|\right].\label{eq:fockexact}
\end{equation}
This time it is the anomaly at $k=k_F$ that we wish to capture with DRP. This is a single-chain example who's real-space expression is given by Eq.~\eqref{eq:chainfin2},
\begin{eqnarray}
\Sigma^{({\rm F})} (r,\tau)&=&-\sum_s\frac{e^{is(k_Fr-\frac{\pi}{2})}}{\lambda_F r}g^{({\rm 1D})}_{e s}(0)V(r,\tau)\\
&=&-\sum_s\frac{e^{is(k_Fr-\frac{\pi}{2})}}{2\pi\lambda_F r}\frac{1}{i sr-v_F\tau}V(r)\delta(\tau),\nonumber\\
\end{eqnarray}
where we have used the fact the Coulomb interaction is instantaneous, i.e. the speed of light is much larger than the Fermi velocity. In frequency space, we have
\begin{eqnarray}
\Sigma^{({\rm F})} (r,\omega)&=&-\frac{\sin(k_Fr-\frac{\pi}{2})}{\pi\lambda_F r^2}V(r).
\end{eqnarray}
Using Eqs.~\eqref{eq:Ft1} and \eqref{eq:Ftfinal}, we get the Fourier transform
\begin{eqnarray}
\Sigma^{({\rm F})} (k,\omega)&=&\frac{4}{\lambda_F}\sqrt{\frac{\pi}{2}}\int_0^\infty dr\cos\left(k_Fr\right)V(r)\frac{J_{1/2}(kr)}{(kr)^{1/2}}\nonumber\\
\\
&=&\frac{4 e^2}{\epsilon\lambda_Fk}\int_\Lambda^\infty dr\cos\left(k_Fr\right)\frac{\sin(kr)}{r^2},
\end{eqnarray}
where we have once again regularized the integral with a short-distance cutoff $\Lambda$. Evaluating the radial integral gives
\begin{eqnarray}
\Sigma^{({\rm F})} (k,\omega)&=&\frac{2 e^2}{\epsilon\lambda_Fk}\bigg[(k-k_F)\mathcal{C}[(k-k_F)r]\nonumber\\
&&+(k+k_F)\mathcal{C}[(k+k_F)r]\nonumber\\
&&-2\frac{\cos(k_Fr)\sin(kr)}{r}\bigg]_\Lambda^\infty.
\end{eqnarray}
Expanding about $k=k_F$ gives
\be
\Sigma^{({\rm F})} (k,\omega)=-\frac{e^2}{\epsilon\pi}(k-k_F)\ln|k-k_F|+{\rm const.},
\ee
which is precisely the leading term of Eq.~\eqref{eq:fockexact}.
\section{Conclusion} 
The above examples involved no computations further than a Fourier transform. For higher-order diagrams the computations are more involved, owing to the intermediate one-dimensional integrals. However, the equivalence we have established between $D$-dimensional diagrammatics and one-dimensional diagrammatics allows one to use the full machinery of 1D analytics to solve higher-dimensional problems. If we can assume that the relevant class of diagrams for a $D$-dimensional problem falls into the category of dimensionally-reducible diagrams (as given by the assumptions in Sec.~\ref{sec:assump}), then we can use bosonization to solve the original problem. Specifically, by computing the $1D$ correlators of left and right movers for the Tomonaga-Luttinger model with interactions given by the $1D$ version of the original interaction, (i.e. $V(|\b{r}|;\tau)\rightarrow V(x,\tau)$), we can evaluate the $D$-dimensional correlators via the equation expressed in Fig.~\ref{fig:dimred}. Up to some slowly-varying analytic terms in momentum space, this correspondence is exact.

The usefulness of this technique has already been demonstrated for a variety of non-trivial systems~\cite{miserev2021a, miserev2022, miserev2023, hutchinson2024, miserev2024, miserev2025}, but these only scratch the surface of strongly correlated phenomena. Several open questions remain. To what extent the DRP can be established for back-scattering interactions is unclear. How it can be modified for systems with alternative types of non-interacting Green's functions, for example, Landau levels or anisotropic Fermi surfaces, is also an interesting question. Perhaps the DRP can even be used to study potential scattering for transport problems. 

\emph{Acknowledgements} --
This work was supported by the Georg H. Endress Foundation and the Swiss National Science Foundation.
\appendix
\section{Angular integrals in $D$ dimensions}
The Fourier transform of a radial function $f(r)$ in $D$ dimensions is
\be
\int d^Dr e^{i\b{k}\cdot\b{r}}f(r)=\int dr r^{D-1}\Omega(r),\label{eq:Ft1}
\ee
where we have identified the angular integral over the $(D-1)$-dimensional sphere of radius $r$, parametrized by angles $\theta_1\ldots\theta_{D-1}$ as
\be
\Omega(r)=\int_0^{2\pi}d\theta_{D-1}\prod_{j=1}^{D-2}\int_0^\pi d\theta_{j}(\sin\theta_{j})^{D-1-j}e^{ikr\cos\theta_1}.\label{eq:Omega}
\ee
The following important identity allows us to evaluate this integral,
\begin{eqnarray}
\int_0^\pi d\theta e^{ikr\cos\theta}\sin^n\theta&=&\int_{-1}^1 dx (1-x^2)^{\frac{n-1}{2}}e^{ikrx}\\
&=&2\int_0^1dx(1-x^2)^{\frac{n-1}{2}}\cos(krx)\nonumber\\
&=&\sqrt{\pi}\Gamma\left(\frac{n+1}{2}\right)\frac{J_{n/2}(kr)}{(kr/2)^{n/2}},\label{eq:Ft}
\end{eqnarray}
where $J_{m}(x)$ is the $m$th-order Bessel function and $\Gamma(x)$ is the gamma function~\cite[\href{https://dlmf.nist.gov/10.9.E4}{(10.9.4)}]{NIST:DLMF}. Taking the limit $k\rightarrow0$ gives another useful identity,
\be
\int_0^\pi d\theta\sin^n\theta=\frac{\sqrt{\pi}}{\Gamma(n/2+1)}\Gamma\left(\frac{n+1}{2}\right),\label{eq:surfacearea}
\ee
which can be used to compute the surface area of a $(D-2)$-sphere,
\be
S=\frac{2\pi^{\frac{D-1}{2}}}{\Gamma\left(\frac{D-1}{2}\right)}.\label{eq:surf}
\ee
Eqs.~\eqref{eq:Ft} and~\eqref{eq:surf} allow us to write~\eqref{eq:Omega} as
\begin{eqnarray}
\Omega(r)&=&S\int_0^\pi d\theta_1e^{ik\cos\theta_1}(\sin\theta_1)^{D-2}\\
&=&(2\pi)^{D/2}\frac{J_{D/2-1}(kr)}{(kr)^{D/2-1}}.\label{eq:Ftfinal}
\end{eqnarray}
\section{Stationary phase approximation in $D$-dimensions}\label{sec:statphase}
We are interested in evaluating angular integrals that make up a chain, which take the form $\Omega\equiv\int d\Omega f(\b{r},\b{r}')$, where $d\Omega$ is the surface element of a $D$-dimensional sphere of radius $r$. The function $f$ has an oscillatory part coming from the Green's function (here we consider a forward-scattering interaction), which takes the form ${\rm exp}[is(k_F|\b{r}-\b{r}'|-\vartheta)]$.
There are two stationary points: when $\b{r}$ is parallel and anti-parallel to $\b{r}'$. Taking the angle between these vectors to be $\theta_1$, we can expand the phase of the exponential to second order about each of these angles,
\begin{eqnarray}
\phi^0(\theta_1)&\approx& s\bigg(k_F|r'-r|-\vartheta+\frac{k_Frr'}{2|r'-r|}\theta_{1}^2\bigg)\\
\phi^\pi(\theta_1)&\approx& s\bigg(k_F|r'+r|-\vartheta-\frac{k_Frr'}{2|r'+r|}(\theta_{1}-\pi)^2\bigg).\nonumber\\
\end{eqnarray}

As with the Fourier transform written in Eq.~\eqref{eq:Omega}, the surface integral of the function $f(\b{r},\b{r}')$ is
\begin{eqnarray}
\Omega&=&\int_0^{2\pi}d\theta_{D-1}\prod_{i=1}^{D-2}\int_0^\pi d\theta_{i}(\sin\theta_{i})^{D-1-i}f(\b{r},\b{r}').\nonumber\\
\end{eqnarray}
Since $f(\b{r},\b{r}')$ only depends on a single angle $\theta_1$, we can integrate the other angles, producing the surface area $S$ of a $(D-2)$-sphere given by Eq.~\eqref{eq:surf}. 
This leaves one angular integration,
\begin{eqnarray}
\Omega&=&S\int_0^\pi d\theta_{1}(\sin\theta_{1})^{D-2}f(\b{r}-\b{r}').\nonumber\\
&\approx&S\int_0^\pi d\theta_{1}[\theta_1^{D-2}e^{i\phi^0(\theta_1)}+(\pi-\theta_1)^{D-2}e^{i\phi^\pi(\theta_1)}]\nonumber\\
&\approx&S\int_0^\infty d\theta_{1}\theta_1^{D-2}[e^{i\phi^0(\theta_1)}+e^{i\phi^\pi(\pi-\theta_1)}].\nonumber\\
\end{eqnarray}
In the second line, we expanded the integrand around the stationary points, and in the third we extended the integration range to infinity, which adds a negligible contribution to the integrals since the the integrands oscillate rapidly away from these points. Using a change of variables $z_0\equiv-i\frac{sk_Frr'}{2|r'-r|}\theta_1^2$ and  $z_\pi\equiv i\frac{sk_Frr'}{2|r'+r|}\theta_1^2$, allows us to write $\Omega$ as
\begin{eqnarray}
\Omega&=&\frac{S}{2}\left(\frac{i2s|r'-r|}{k_Frr'}\right)^{\frac{D-1}{2}}\int_0^\infty dz_0z_0^{\frac{D-3}{2}}e^{-z_0}e^{i\phi^0(0)}\nonumber\\
&&+\frac{S}{2}\left(\frac{-i2s|r'+r|}{k_Frr'}\right)^{\frac{D-1}{2}}\int_0^\infty dz_\pi z_\pi^{\frac{D-3}{2}}e^{-z_\pi}e^{i\phi^\pi(\pi)}.\nonumber\\
\end{eqnarray}
We recognize these integrals as the defining representation of the gamma function $\Gamma(x)=\int_0^\infty dz z^{x-1}e^{-z}$, which cancels with the gamma function in the surface area factor~\eqref{eq:surf},
\begin{eqnarray}
\Omega&=&\left(\frac{2\pi|r'-r|}{k_Frr'}\right)^{\frac{D-1}{2}}e^{is\frac{\pi}{4}(D-1)}e^{is(k_F|r'-r|-\vartheta)}\nonumber\\
&&+\left(\frac{2\pi|r'+r|}{k_Frr'}\right)^{\frac{D-1}{2}}e^{-is\frac{\pi}{4}(D-1)}e^{is(k_F|r'+r|-\vartheta)}.\nonumber\\
\end{eqnarray}
Introducing the relative index $t=\pm1$, which gives the relative sign between $r$ and $r'$ (or the relative orientation of the two corresponding rotors) and recalling that $\vartheta=\frac{\pi}{4}(D-1)$, we finally have
\be
\Omega=\sum_{t=\pm1}\left(\frac{\lambda_F|r'-tr|}{rr'}\right)^{\frac{D-1}{2}}e^{is[k_F|r'-tr|-\frac{\pi}{4}(D-1)(1-t)]}.\label{eq:Omegafinal}
\ee


\bibliographystyle{apsrev4-1}
\bibliography{bibdimred}
\end{document}